\documentclass{elsart}

\usepackage{amssymb}
\usepackage{graphicx}

\def\BibTeX{{\rm B\kern-.05em{\sc i\kern-.025em b}\kern-.08em
    T\kern-.1667em\lower.7ex\hbox{E}\kern-.125emX}}

\begin{document}

\begin{frontmatter}

\title{Is a hyperchaotic attractor superposition of two multifractals?}

\author[label1]{K. P. Harikrishnan\corauthref{cor1}}
\ead{$kp.hk05@gmail.com$}
\author[label2]{R. Misra}
\ead{$rmisra@iucaa.in$}
\author[label3]{G. Ambika}
\ead{$g.ambika@iiserpune.ac.in$}

\corauth[cor1]{Corresponding author: Address: Department of Physics, The Cochin College,  Cochin-682002, India; Phone No.0484-22224954;  Fax No: 91-22224954.} 

\address[label1]{Department of Physics, The Cochin College, Cochin-682002, India}
\address[label2]{Inter University Centre for Astronomy and Astrophysics, Pune-411007, India}
\address[label3]{Indian Institute of Science Education and Research, Pune-411008, India}

\begin{abstract}
In the context of chaotic dynamical systems with exponential divergence of nearby trajectories in phase 
space, hyperchaos is defined as a state where there is divergence or stretching in at least two directions  
during the evolution of the system. Hence the detection and characterization of a hyperchaotic attractor is 
usually done using the spectrum of Lyapunov Exponents (LEs) that measure this rate of divergence along each 
direction.  Though hyperchaos arise in different dynamical situations and find 
several practical applications, a proper understanding of the geometric structure of a hyperchaotic 
attractor still remains an unsolved problem. In this paper, we present strong numerical evidence to 
suggest that the geometric structure of a hyperchaotic attractor can be characterized using a multifractal 
spectrum with two superimposed components. In other words, apart from developing an extra positive LE,  
there is also a structural change as a chaotic attractor makes a transition to the hyperchaotic 
phase and the attractor changes from a simple multifractal to a dual multifractal, equivalent to two 
inter-mingled multifractals. We argue that a cross-over behavior in the scaling region for computing the 
correlation dimension is a manifestation of such a structure. 
In order to support this claim, we present an illustrative example of a synthetically generated set of points 
in the unit interval (a Cantor set with a variable iteration scheme)   
displaying dual multifractal spectrum. Our results are also used to develop a general scheme to generate both 
hyperchaotic as well as high dimensional chaotic attractors by coupling two low dimensional chaotic 
attractors and tuning a time scale parameter.    
\end{abstract}

\begin{keyword}

Hyperchaotic attractor\sep Multifractals \sep Time series analysis

\end{keyword}

\end{frontmatter}

\section{Introduction}
A dynamical system is considered to be chaotic if it shows the property of sensitive 
dependence on initial conditions. For such systems, two nearby trajectories diverge exponentially in 
time during 
the evolution of the system, indicating that one of the LEs is positive. Hyperchaos is formally 
defined as a state where there is divergence in at least two directions as the system evolves. 
Hyperchaotic attractors are thus characterized by at least two positive LEs and 
are considered to be much more complex in terms of topological structure and dynamics 
compared to low dimensional chaotic attractors. In the last two decades, hyperchaotic systems 
have attracted increasing attention from various scientific and engineering communities due to a 
large number of practical applications. These include secure communication and 
cryptography \cite {sho,jju}, synchronization studies using electro-optic devices \cite {lar,goe} 
and as a model for chemical reaction chains \cite {bai}. In all these applications, the 
complexity of the underlying attractor has a major role to play.

Though the concept of hyperchaos was introduced many years ago \cite {ros}, a 
systematic understanding of the topological and fractal structure of the attractors generated 
from the hyperchaotic systems is lacking till date. Studies in this direction have 
been very few except a series of papers on a system of unidirectionally coupled oscillators 
\cite {kap1,kap2,kap3} in which the authors have discussed many aspects of the structure and 
transition to hyperchaos in the model, including dual scaling regions in the hyperchaotic phase.  

Hyperchaotic attractors generated by continuous systems are, in general, higher dimensional 
with the fractal dimension $D_0 > 3$ and 
trajectories diverging in at least two directions as the system evolves in time. Hence the 
detection of hyperchaos is generally done using the LEs with the transition to hyperchaos 
marked by the crossing of the second largest LE above zero. One of our aims in this paper is 
to try and get a more quantitative information regarding the structure of the hyperchaotic 
attractor in terms of the spectra of dimensions and use this information to  detect the 
transition to hyperchaos. 

Recently, we have done a detailed 
dimensional analysis \cite {kph1,kph2} of several standard hyperchaotic models and have 
established some results which are common to all these systems. We have applied a  
modified box counting scheme and have obtained an improved scaling region for computing the 
fractal dimension of the system.  Also, we have shown 
that the topological structure of the underlying attractor changes suddenly as the system makes 
a transition from chaos to hyperchaos and there is a cross-over behavior in the scaling of the 
correlation dimension $D_2$ resulting in two different scaling regions in the hyperchaotic 
phase. Here we investigate this cross-over behavior in more detail numerically and show that 
we can derive the whole spectrum of $D_q$ values corresponding to the two different scaling 
regions. We consider this result as a consequence of the fact that the geometric structure of a 
hyperchaotic attractor is equivalent to that of two inter-mingled multifractals and the 
cross-over property is a manifestation of this structure. 
In other words, the overall fractal structure of a hyperchaotic attractor 
can be characterized by two superposed $f(\alpha)$ spectrums. 

It should be noted that the \emph{multiscales} exhibited by multifractals have recently become 
an interesting area of research and have been discussed in various contexts. For example, the 
importance of multiscale multifractal analysis (MMA) has been demonstrated in the study of 
human heart rate variability time series \cite {gie}, where the multifractal properties of the 
measured signal depends on the time scale of fluctuations or the frequency band. Also, 
multiscale multifractal intermittent turbulence in space plasmas has been investigated in the 
time series of velocities of solar wind plasma \cite {mac}.   
In order to convince the reader that a dual multifractal structure can be realized in practice, 
we generate a Cantor set using variable iteration scheme which displays dual slopes in the scaling 
region. Finally, the specific information regarding the structure of the hyperchaotic attractor provides 
us the possibility of generating hyperchaos by coupling two chaotic attractors, a result already 
shown in the literature \cite {gras,lu}. Here we present a general scheme for this to get both 
hyperchaos and high dimensional chaos by varying a control parameter.  

Our paper is organized as follows: In the next section, we present a brief summary of the standard 
multifractal approach for a point set. In \S 3, we discuss the details of numerical 
computations of the multifractal spectrum to show how the structure of a hyperchaotic attractor 
varies from that of an ordinary chaotic attractor. In order to validate our arguments regarding 
the structure of the hyperchaotic attractor, we present an example of a system having 
analogous structure in \S 4 which is a synthetically generated Cantor set 
using a specific iterative scheme. The details regarding the generation of hyperchaos based on our 
numerical results are discussed in \S 5. The paper is concluded in \S 6. 

\begin{figure}
\begin{center}
\includegraphics*[width=16cm]{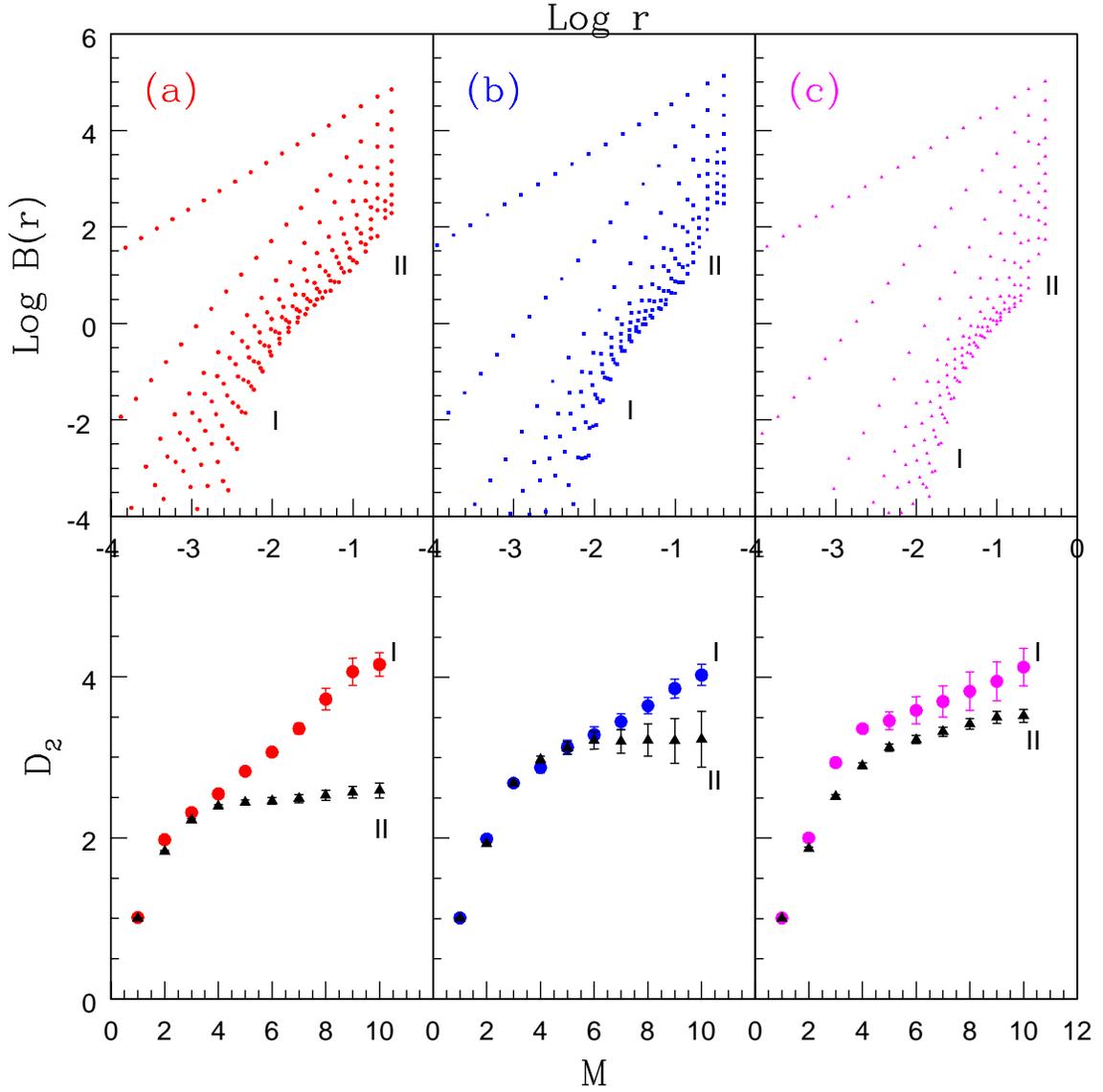}
\end{center}
\caption{Correlation dimension computed using modified box counting scheme for three 
different hyperchaotic time series. The top panel shows the scaling region for computing $D_2$ for 
(a) the Chen hyperchaotic flow (b) the M-G system and (c) the Ikeda system. Each curve is for 
an embedding dimension $M$ that varies from 1 to 10. Two scaling regions I and II are evident 
in all the three cases for $M > 3$.  The bottom panel shows the variation of 
$D_2$ for both scaling regions with $M$ for the three cases.}
\label{f.1}
\end{figure}

\section{Mathematical preliminaries}  

It is well known that, unlike ideal fractals, real world systems and limited point sets exhibit 
self similarity only over a finite range of scales \cite {ber}. Thus in the present case, 
statistical self similarity and hence the multifractal behavior changes between two finite 
range of scales. Multifractality is commonly related to a probability measure that can have 
different fractal dimensions on different parts of the support of this measure. Many authors 
have discussed the standard multifractal approach in detail \cite {hil,pal,hal,gra1} and we 
briefly summarise the main results below for a point set (such as, an attractor generated by a 
chaotic system). 

Let the attractor be partitioned into $M$ dimensional cubes of side $r$, with $N(r)$ being the 
number of cubes required to cover the attractor. If $p_i(r)$ is the probability that the 
trajectory passes through the $i^{th}$ cube, then $p_i(r) = N_i/N_p$, where $N_i$ is the number of 
points in the $i^{th}$ cube and $N_p$ the total number of points on the attractor. We now assume 
that $p_i(r)$ satisfies a scaling relation
\begin{equation}
p_i(r) \propto r^{\alpha_i}
  \label{eq:1}
\end{equation}
where $\alpha_i$ is the scaling index for the $i^{th}$ cube. We now ask how many cubes have the 
same scaling index $\alpha_i$ or have scaling index within $\alpha$ and $\alpha + d\alpha$ (if 
$\alpha$ is assumed to vary continuously). Let this number, say $g(\alpha) d\alpha$, scales 
with $r$ as 
\begin{equation}
g(\alpha) \propto r^{-f(\alpha)}
  \label{eq:2}
\end{equation}
where $f(\alpha)$ is a characteristic exponent. Obviously, $f(\alpha)$ behaves as a dimension and 
can be interpreted as the fractal dimension for the set of points with scaling index $\alpha$. 
This also implies that the attractor can be characterized by a spectrum of dimensions normally 
denoted by $D_q$ (where $q$ can, in principle, vary from $-\infty$ to $\infty$) \cite {hen}, 
that can be related to $f(\alpha)$ through a Legendre transformation \cite {atm}. The plot of 
$f(\alpha)$ as a function of $\alpha$ gives a one hump curve with maximum corresponding to 
$D_0$, the simple box counting dimension of the attractor.

Note that, in the above arguement, the scaling exponent $\alpha$ measures how fast the number of 
points within a box decreases as $r$ is reduced. 
It therefore measures \emph{the strength of a singularity} for $r \rightarrow 0$. For a 
realistic attractor, with limited number of data points, the limit $r \rightarrow 0$ is not 
accessible and hence one chooses a suitable scaling region for $r$ to compute $\alpha$ and 
$f(\alpha)$.
This is where a hyperchaotic attractor becomes different from an ordinary chaotic attractor, as 
per our numerical results. We find that, to characterize the multifractal structure of a 
hyperchaotic attractor, two seperate scaling regions are to be considered indicative of the 
presence of two underlying multifractals. The detailed numerical results are presented in 
the next section.

\begin{figure}
\begin{center}
\includegraphics*[width=16cm]{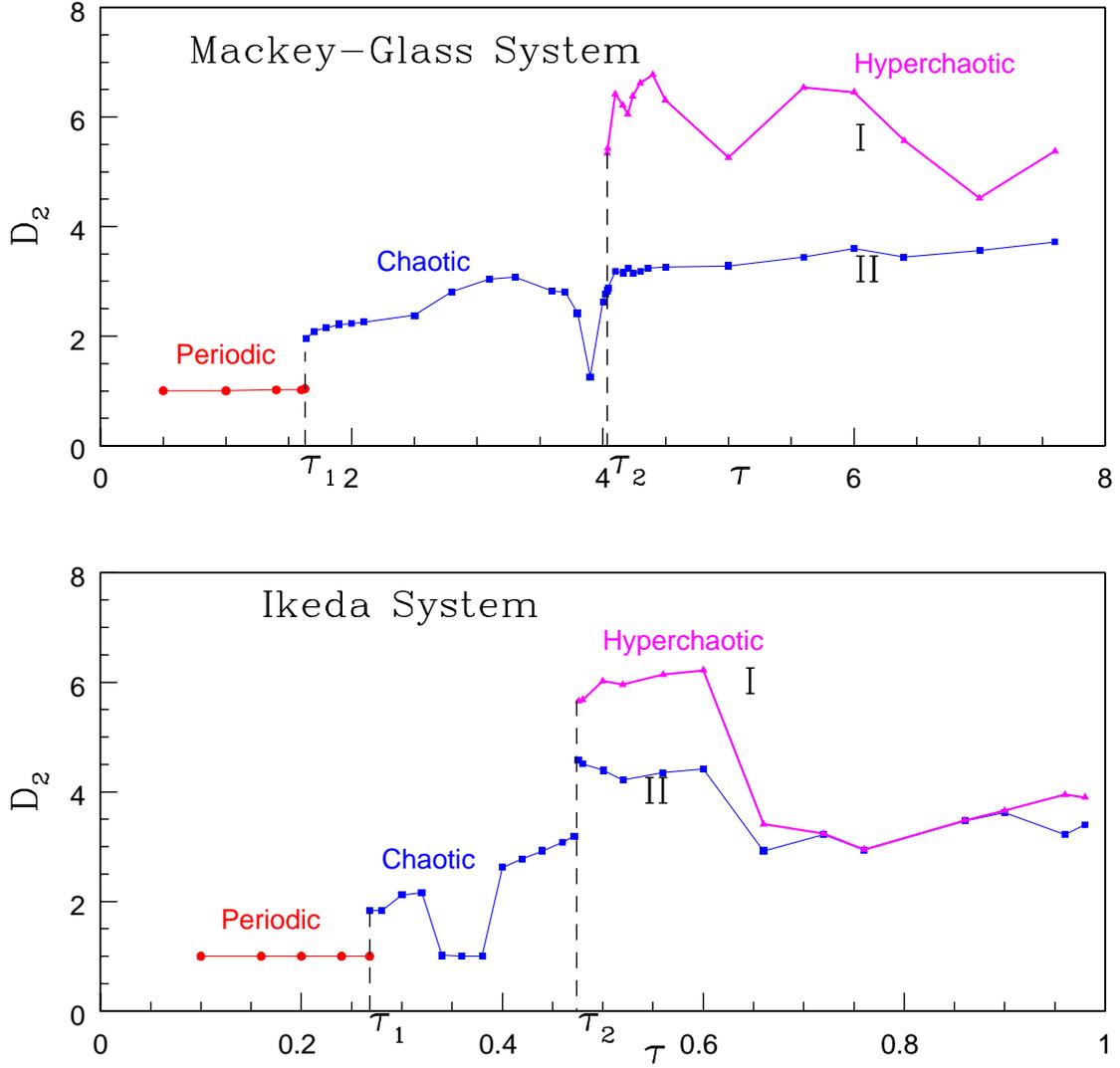}
\end{center}
\caption{Transition from periodic to chaotic and hyperchaotic phase represented as a function of 
the time delay parameter $\tau$ for two standard time delayed systems, with $D_2$ as a 
quantifying measure to distinguish between the three phases. Note the presence of two $D_2$ values 
in the hyperchaotic phase corresponding to two scaling regions, indicated as I and II.}
\label{f.2}
\end{figure}

\begin{figure}
\begin{center}
\includegraphics*[width=16cm]{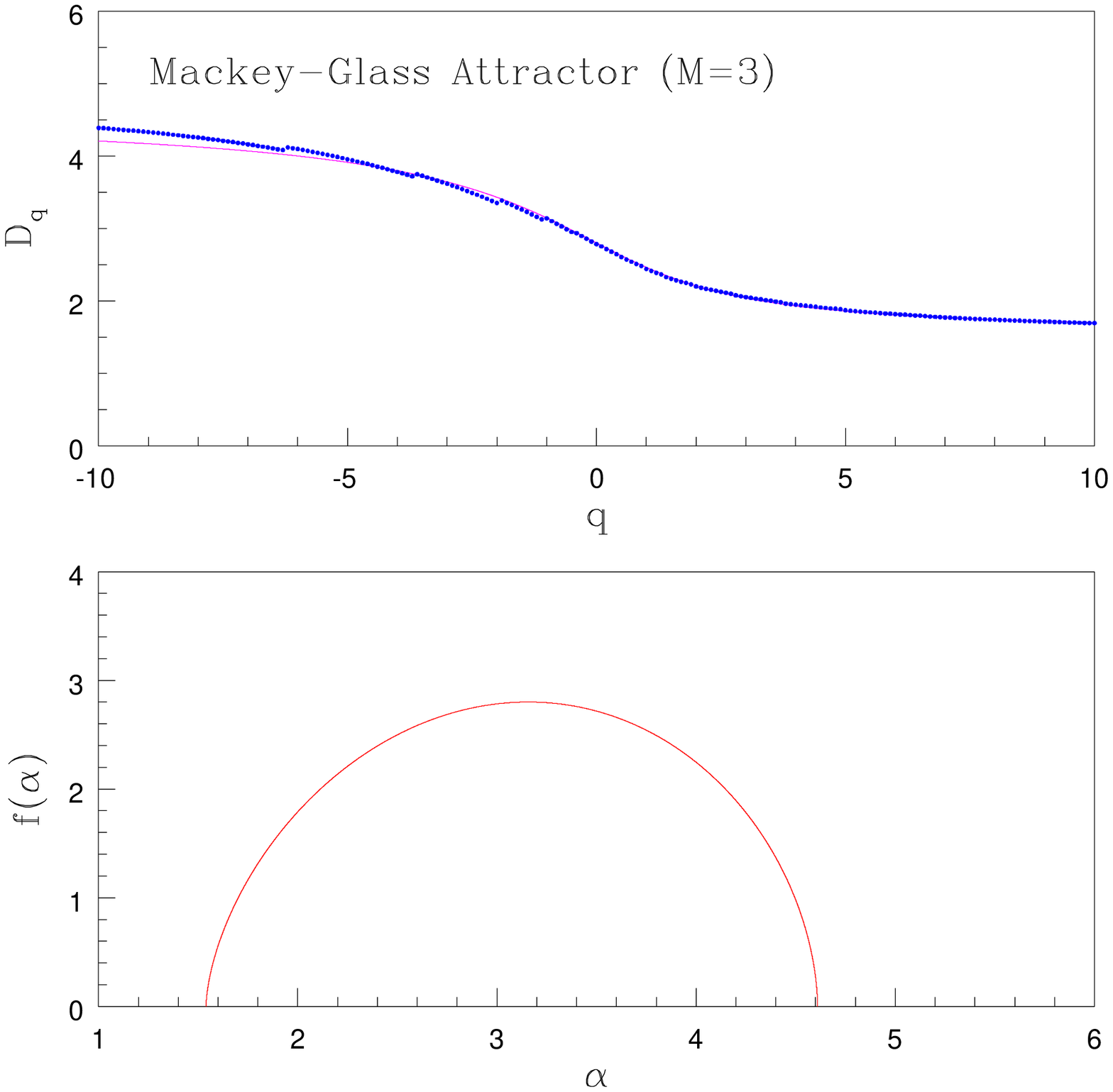}
\end{center}
\caption{Top panel shows the $D_q$ values (points) and its best fit curve 
(continuous line) of the M-G attractor for $\tau = 3.70$ (in the chaotic phase), calculated 
from a time series consisting of $2 \times 10^5$ data points with $M = 3$. The lower panel 
shows the $f(\alpha)$ spectrum computed from the best fit curve.}
\label{f.3}
\end{figure}

\section{Hyperchaotic attractor as a dual multifractal}
Before going into the computation of the multifractal spectrum, we discuss very briefly our 
results on $D_2$ obtained using the modified box counting scheme \cite {kph1,kph2}, where the 
scaling region for computing $D_2$ is fixed algorithmically. The attractor is covered using 
$M$-dimensional cubes of size $r$. The probability $p_i$ that the trajectory passes through 
the $i^{th}$ cube is computed by taking an ensemble average of the number of points falling 
in the $i^{th}$ cube. This modifies the equation for computing the weighted box counting sum 
$B(r)$ as:
\begin{equation}
B(r) = {{1} \over {N_p^2}}[{\sum_i m_i^2} - N_p]
  \label{eq:3}
\end{equation}
where $N_p$ is the total number of points on the attractor and $m_i$ is the number of points 
falling in the $i^{th}$ box. The correlation dimension is then calculated as the scaling 
index of the variation of $B(r)$ with $r$ as: 
\begin{equation}
 D_{2} \equiv \lim_{r \rightarrow 0} {\log B(r)}/{\log (r)}
  \label{eq:4}
\end{equation}

For the present study, we use time series from three standard 
hyperchaotic systems, namely, the Chen hyperchaotic flow \cite {che}, the 
Mackey-Glass (M-G) time delayed system \cite {mcm} and the Ikeda time delayed system 
\cite {ike}. For the hyperchaotic flow, we fix the parameters as studied in detail in 
\cite {kph1} $(a=35, b=4.9, c=25,d=5,e=35,k=100)$ to generate the hyperchaotic time series. 
For M-G and Ikeda systems, we use the time delay $\tau$ as the control parameter with the 
other parameters fixed as $\beta=2, \gamma=1, n=10$ for M-G and $a=5, m=20$ for Ikeda 
respectively. We have studied the transition to hyperchaos in these two time delayed systems in 
detail \cite {kph2} and here we choose $\tau=6.40$ for M-G and $\tau=0.56$ for Ikeda for 
generating the hyperchaotic time series.

In Fig.~\ref{f.1} (top panel), we show the scaling region computed by our modified scheme 
for the above three systems in the hyperchaotic 
phase. Here, the weighted box counting sum $B(r)$ \cite {kph2} is plotted against box size $r$. 
Each curve corresponds to an embedding dimension $M$, which varies from $1$ to $10$. Two 
scaling regions (denoted I and II) are evident in all cases, for $M \geq 4$. The $D_2$ values 
computed from the two scaling regions as a function of $M$ are also shown in Fig.~\ref{f.1} 
(bottom panel). Based on our numerical results, the transition from chaos to hyperchaos can be 
identified with the sudden appearance of a second scaling region with a distinct value of $D_2$. 
This is shown for two standard time delayed systems in Fig.~\ref{f.2}, taking the time delay 
$\tau$ as the control parameter. 
We argue that this result is an indication of a structural change in the 
underlying chaotic attractor as the system makes a transition to the hyperchaotic phase at a 
critical value of the control parameter. The attractor changes from a simple multifractal to a 
dual multifractal. To convince this, we now compute the entire $D_q$ and $f(\alpha)$ spectra 
corresponding to the two scaling regions seperately, which is the focus of the present paper. 
Note that the term ``dual multifractal''has already been used in a different context in the 
literature by Roux and Jenson \cite {roux} while referring to the use of two related functions 
to calculate the multifractal spectra by using Cantor set as example. It should not be confused 
with our terminology since we only use the term to indicate the structure of two 
inter-mingled multifractals.   

\begin{figure}
\begin{center}
\includegraphics*[width=16cm]{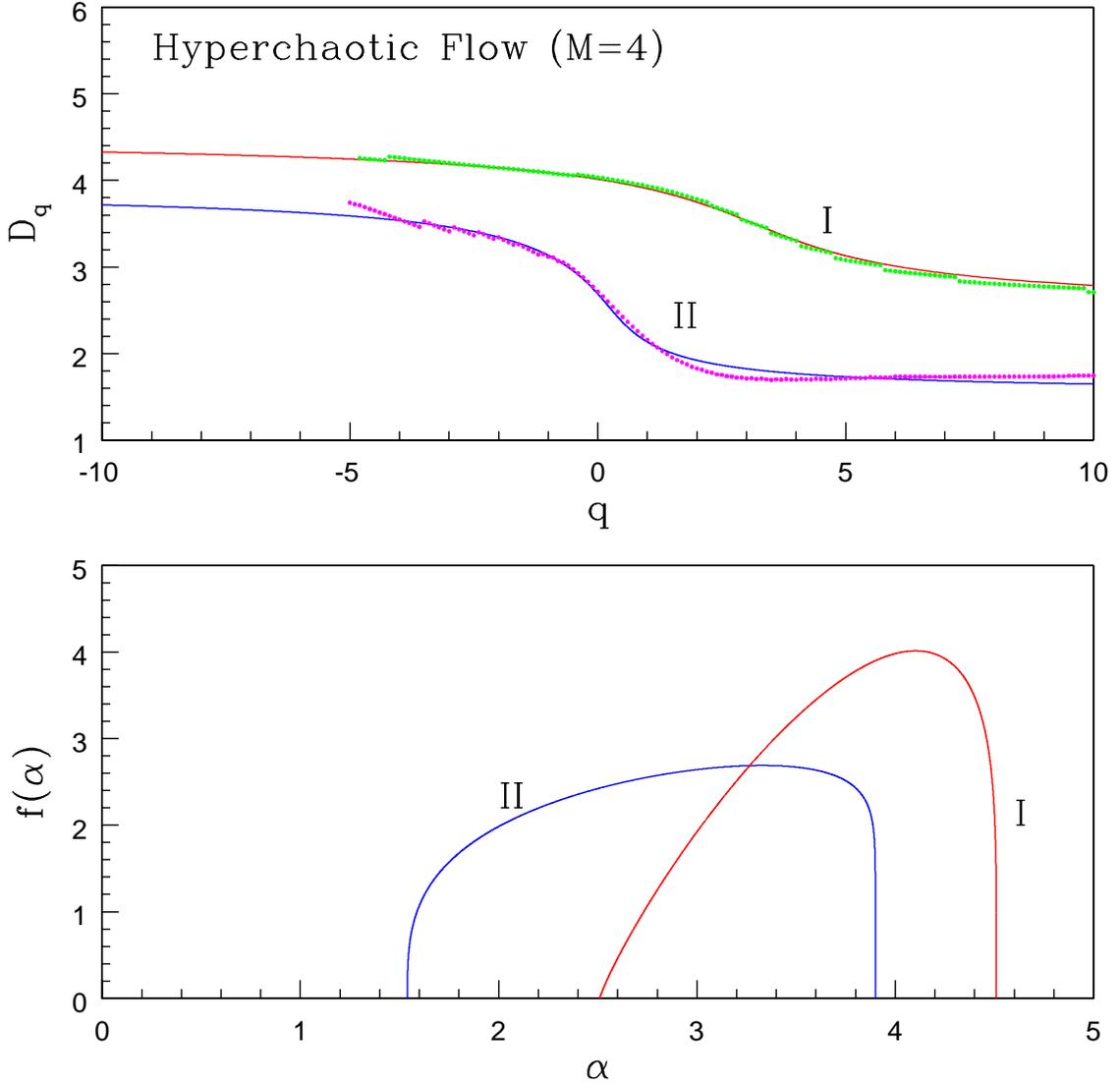}
\end{center}
\caption{Top panel shows the $D_q$ values (points) and the best fit curve 
(line) for the two scaling regions of the Chen hyperchaotic flow in the hyperchaotic phase 
with $M = 4$. The lower panel shows the corresponding $f(\alpha)$ spectra computed from 
the best fit curves.}
\label{f.4}
\end{figure}

\begin{figure}
\begin{center}
\includegraphics*[width=16cm]{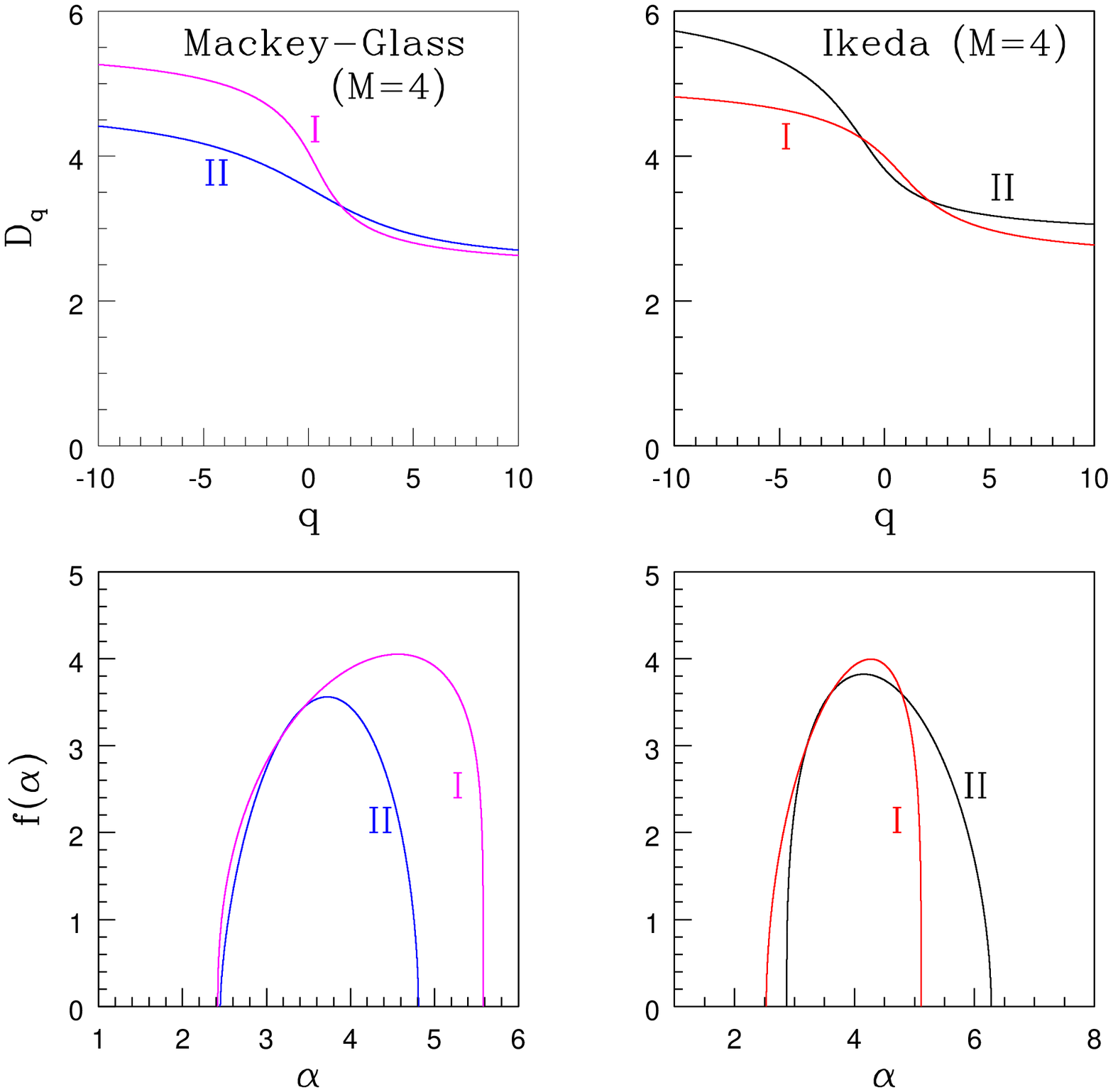}
\end{center}
\caption{The best fit $D_q$ curves (top panel) for the two time delay systems and their corresponding 
$f(\alpha)$ spectra (bottom panel) in the hyperchaotic phase.}
\label{f.5}
\end{figure}

To compute the $f(\alpha)$ spectrum, we use the automated algorithmic scheme proposed by us 
a few years back \cite {kph4}. A brief discussion of the scheme is given 
below which is based on the Grassberger-Procaccia (G-P) algorithm \cite {gra2}. 
As the first step, the spectrum of generalised dimensions $D_q$ is computed from the 
time series using the equation
\begin{equation}
    D_{q} \equiv   \frac {1}{q-1} \; \lim_{r \rightarrow 0} \frac{\log \; C_q (r)}{\log \; r}
    \label{eq:5}
\end{equation} 
where $C_q (r)$ are the generalized correlation sum. This is done by choosing the scaling 
region algorithmically as discussed earlier \cite {kph4}. 
We then use an entirely different algorithmic approach for the computation of the smooth 
profile of the $f(\alpha)$ spectrum. The $f(\alpha)$ function is a single valued function 
between $\alpha_{max}$ and $\alpha_{min}$ and also has to satisfy several other 
conditions, such as, it has a single maximum and $f(\alpha_{max}) = f(\alpha_{min}) = 0$. 
A simple function that can satisfy all the necessary conditions is
\begin{equation}
   f(\alpha) = A(\alpha - \alpha_{min})^{\gamma_1}(\alpha_{max} - \alpha)^{\gamma_2}
   \label{eq:6}
\end{equation} 
where $A$, $\gamma_1$, $\gamma_2$, $\alpha_{min}$ and $\alpha_{max}$ are a 
set of parameters characterizing a particular $f(\alpha)$ curve. It can be shown 
\cite {kph4} that only four of these parameters are independent and any general 
$f(\alpha)$ curve can be fixed by four independent parameters. Moreover, by imposing 
the conditions on the $f(\alpha)$ curve, it can also be shown that 
$0 < \gamma_1, \gamma_2 < 1$.

The scheme first takes $\alpha_1 (\equiv D_1), \alpha_{min} (\equiv D_{\infty})$ 
and $\alpha_{max} (\equiv D_{-\infty})$ as input parameters from the 
computed $D_q$ values and choosing an initial value for $\gamma_1$ in the 
range $[0,1]$, the parameters $\gamma_2$ and $A$ are calculated. The 
$f(\alpha)$ curve is then computed in the range $[\alpha_{min}, \alpha_{max}]$. 
From this, a smooth $D_q$ versus $q$ curve can be obtained by inverting using 
the Legendre transformation equations, which is then fitted to the $D_q$ spectrum derived 
from the time series. The parameter values are changed continuously until the $D_q$ 
curve matches with the $D_q$ spectrum from the time series and the statistically best fit 
$D_q$ curve is chosen. From this, the final $f(\alpha)$ curve can be evaluated.

To illustrate the scheme, we first compute the $D_q$ values and the associated $f(\alpha)$ 
spectrum for the M-G attractor in the chaotic phase for $\tau = 3.70$ with embedding 
dimension $M = 3$. The results are shown in Fig.~\ref{f.3}. The $D_q$ values (points)and 
the best fit curve (continuous line) are shown in the upper panel while the associated 
$f(\alpha)$ spectrum computed from the best fit curve is shown in the lower panel.

We now apply the scheme to the hyperchaotic attractors and compute the $D_q$ and the 
$f(\alpha)$ spectrum for the two scaling regions seperately, by fixing $r_{min}^I (r_{min}^{II})$ and 
$r_{max}^I (r_{max}^{II})$ for scaling region I (II). The embedding dimension used is $M = 4$ and 
the number of data points in the time series is $3 \times 10^5$ in all cases. 
Fig.~\ref{f.4} shows the results of computations of the Chen hyperchaotic flow while 
Fig.~\ref{f.5} shows that of the two time delayed systems in the hyperchaotic phase. In both 
cases, the $D_q$ curves corresponding to the two scaling regions are given in the upper panel 
and the associated $f(\alpha)$ spectra are given in the lower panel. Our results indicate that 
the fractal structure of a hyperchaotic attractor is much more complex compared to that of an 
ordinary chaotic attractor having a structure equivalent to a superposition of two multifractals.

\section{A Cantor set with dual multifractal structure}
Here we show that it is possible to generate synthetically a set of points in the unit 
interval $[0,1]$ whose geometric structure can be represented by a dual multifractal. 
We call this set a \emph {2-level 2-scale Cantor set}. The construction of this Cantor 
set is done as follows:\\
For the usual Cantor set, the measure over a unit interval $[0,1]$ is divided into two parts 
with probabilities $p_1$ and $p_2$ (where $p_2 = 1 - p_1$) and assigned to two fractional 
length scales $l_1$ and $l_2$ respectively in the first step. This process is repeated to 
each of the lengths $l_1$ and $l_2$ in the second step. By continuing  this process $n$ 
times (with $n \rightarrow \infty$), one gets the 2 - scale Cantor set which is a 
multifractal. 

\begin{figure}
\begin{center}
\includegraphics*[width=16cm]{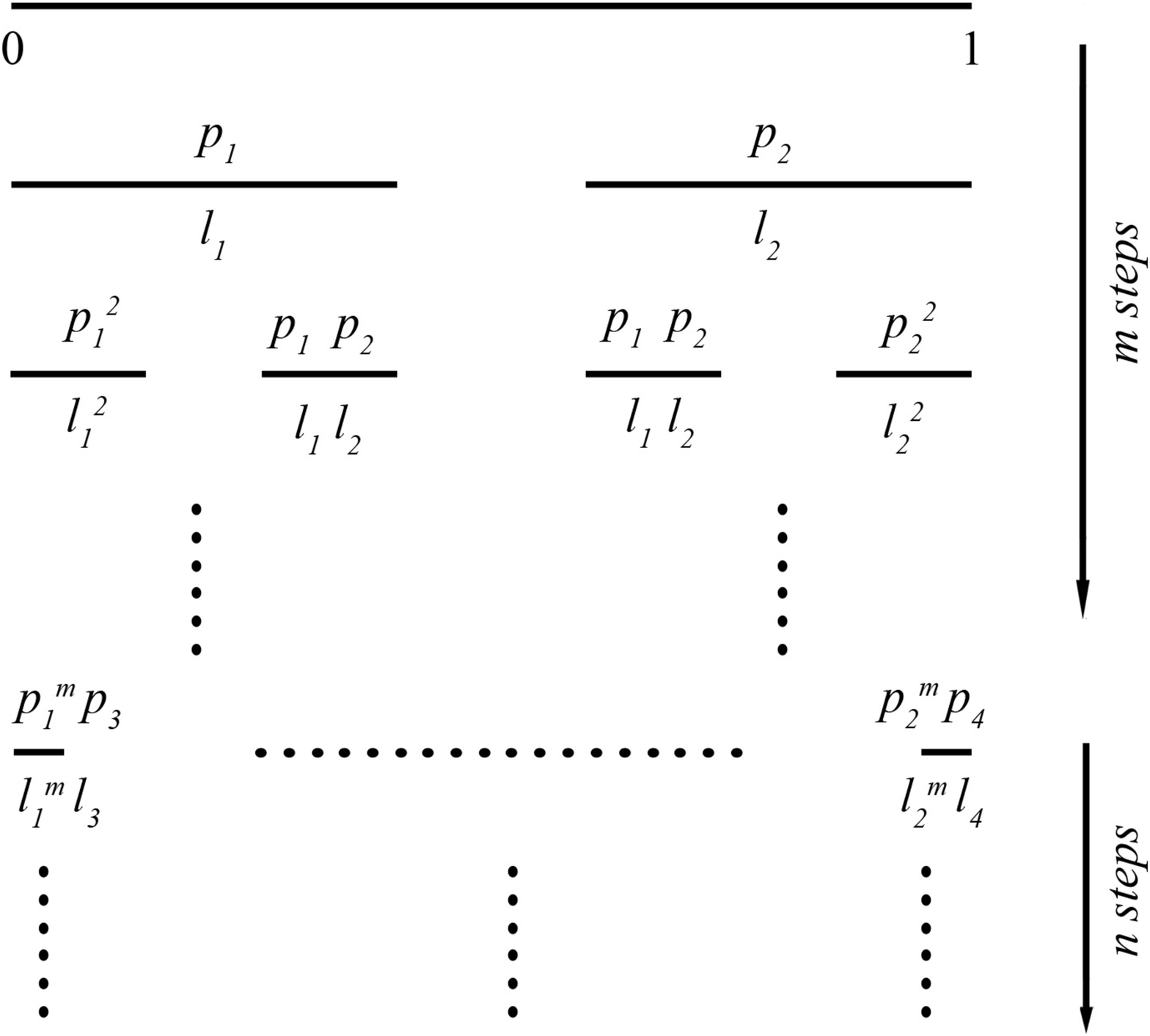}
\end{center}
\caption{Construction of 2 - level 2 - scale Cantor set. A unit interval is divided into 
two fractional lengths $l_1$ and $l_2$ with probabilities of measure $p_1$ and $p_2$ respectively. 
This process is repeated for $m$ steps. For the next $n$ steps, a different set of parameters 
$(l_3, l_4, p_3, p_4)$ is used. The resulting set is the 2- level 2 - scale Cantor set which is a 
superposition of two multifractals for an intermediate range of $m$ values.}
\label{f.6}
\end{figure}

\begin{figure}
\begin{center}
\includegraphics*[width=16cm]{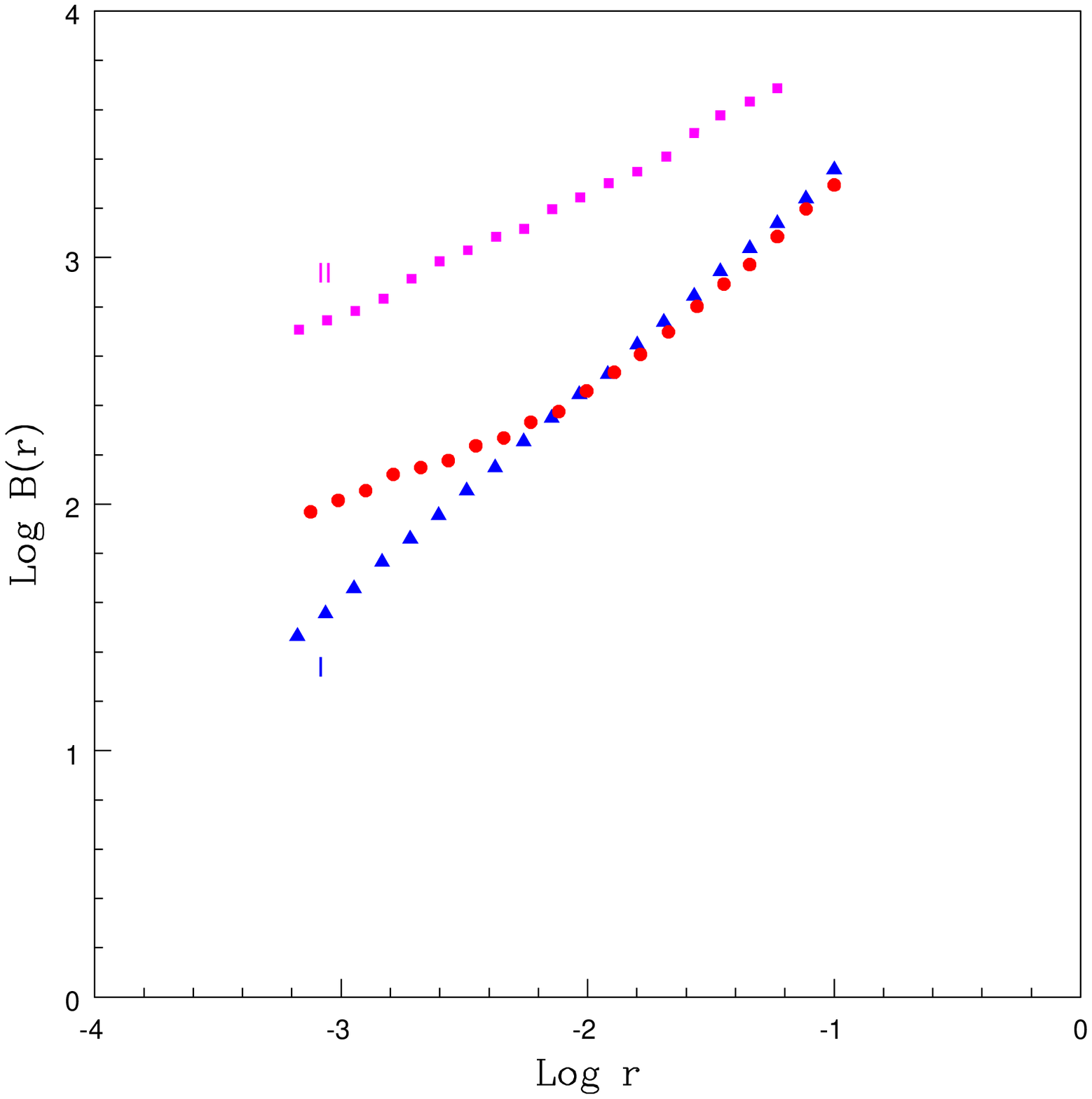}
\end{center}
\caption{Scaling region of the 2 - level 2 - scale Cantor set constructed in the 
previous figure. The solid triangles marked I represent the scaling region for large $m$ 
corresponding to the ordinary 2 - scale Cantor set with parameters in level I while the solid 
squares marked II correspond to small $m$ for the level II Cantor set. The solid circles 
having dual slope represent the scaling region of the resulting set for $m = 100$ which is a 
superposition of the two Cantor sets for levels I and II.}
\label{f.7}
\end{figure}

We now modify this process slightly. The above step is repeated $m$ times where $m$ is a 
finite number which is the \emph {level I} of construction. In the \emph {level II}, we 
change the probabilities $p_1$ and $p_2$ to new values $p_3$ and $p_4$ and the fractional 
length scales from $(l_1,l_2)$ to $(l_3,l_4)$ and continue the procedure for the next $n$ 
steps. The resulting set is the 2 - level 2 - scale Cantor set, whose construction is 
shown in Fig.~\ref{f.6}. 

To implement this numerically, we use the set of parameter values: 
$l_1 = 0.48, l_2 = 0.40, p_1 = 0.56, p_2 = 0.44$,  
$l_3 = 0.62, l_4 = 0.26, p_3 = 0.26, p_4 = 0.74$ so that the fractal dimensions of the sets 
generated by the two levels are widely different. We construct the set using different 
values of $m$ and fixing $n$ to be a large value ($\sim 5000$). We find that if $m$ is 
very small (say, $< 20$), the scaling region for the weighted box counting sum 
$B(r)$ as a function of $r$ has a single slope corresponding to the fractal dimension of 
the Cantor set for level II and for large $m$ ($> 500$), the Cantor set corresponding to 
level I prevails. However, for an intermediate range of $m$ values, the scaling region 
displays two slopes with a smooth transition from one to the other implying that the 
resulting set is a superposition of two Cantor sets involving both levels I and II. 
The above results are shown in Fig.~\ref{f.7} where the intermediate value of $m$ used is $100$. 
Our numerical experiment shows that the 2-level 2-scale Cantor set offers a typical 
geometric construction of a fractal set that displays a multifractal spectrum with two 
superimposed components, analogous to that of a hyperchaotic attractor which occurs in a 
higher dimension. 

\section{Generation of hyperchaos by coupling of two chaotic attractors}
Finally, we show that our numerical result on the structure of a hyperchaotic attractor provides us 
with a possibility for constructing a hyperchaotic system by coupling two chaotic systems. 
There are already a few papers in the literature where the authors propose the construction 
of a hyperchaotic system either by coupling two chaotic systems \cite {gras,lu} or by 
introducing a time delayed transformation \cite {xu}. It has also been suggested that in some 
cases, even the coupling is not required; just an amalgam of two chaotic attractors can 
sometimes become hyperchaotic \cite {peco,tsim}, which strongly supports our results. This 
method is applied in the synchronization studies by multiplexing two chaotic signals \cite {tsim}. 
Here we consider the prospect of generating hyperchaos by coupling any two regular chaotic 
systems and study under what conditions and coupling schemes one can achieve this. 

The coupling of two chaotic systems is more generally employed in the synchronization studies 
\cite {peco} rather than hyperchaos generation where the coupling strength has to be sufficiently 
high. Here we consider two individual chaotic systems evolving at two different time scales 
coupled together. The general scheme that we propose is given by
\begin{eqnarray}
{\bf \vec X} = \tau_{1m} \vec F({\bf \vec X}) + \epsilon_1 \vec H({\bf \vec X, \vec Y})     \nonumber \\
{\bf \vec Y} = \tau_{2m} \vec F({\bf \vec Y}) + \epsilon_2 \vec H({\bf \vec X, \vec Y})
\label{eq:7}
\end{eqnarray} 
Here $({\bf \vec X, \vec Y})$ $\in \mathcal R^{n}$ with $\vec F$ representing the intrinsic dynamics of the two 
systems and $\vec H$ denoting the coupling function. The parameters $\epsilon_1$ and $\epsilon_2$ 
represent the coupling strengths between the two systems and the parameters $\tau_{1m}$ and 
$\tau_{2m}$ are the two time scale parameters indicating that the two systems are evolving at two different 
time scales. Without loss of generality, we can take $\tau_{1m} = 1, \tau_{2m} = \tau_{m}$ and 
$\epsilon_1 = \epsilon_2 = \epsilon$. 

The two individual systems ${\bf \vec X}$ and ${\bf \vec Y}$ can be any two low dimensional chaotic 
systems, identical or different. We have done a detailed numerical analysis taking some standard low 
dimensional chaotic systems, such as, Lorenz, R\"ossler and Ueda as individual systems with 
$\vec F({\bf \vec X})$ and $\vec F({\bf \vec Y})$ identical as well as different. We use the standard 
parameters for the individual systems in the chaotic regime. We have tried both diffusive coupling 
and the linear coupling, the two commonly used coupling schemes. In the former case, the feedback terms  
used for coupling are $\epsilon (\vec y - \vec x)$ and $\epsilon (\vec x - \vec y)$ while in the latter, 
these are $\epsilon \vec y$ and $\epsilon \vec x$. In both cases, the 2-way or mutual 
coupling as well as the 1-way or drive-response coupling have been tested.    
The parameter $\epsilon$ should be sufficiently small $(< 0.1)$ in order to avoid the 
synchronization of the dynamics between the two systems and the amplitude death \cite {peco,kajari}. 
We vary the value of $\epsilon$ in the range $0.02$ to $0.05$. The time step of integration 
$\Delta t$ should be sufficiently small to capture the small scale properties of the hyperchaotic 
attractor and hence we fix the value of $\Delta t = 0.002$ in our numerical simulations.  
Our control parameter is $\tau_m$. We have found that hyperchaos can be generated in all the different 
coupling schemes mentioned above for a range of values of $\tau_m$ depending on the individual 
systems, nature of coupling and strength of coupling. 

To show the above results explicitly, we consider two specific cases. In the first case, we choose 
the diffusive coupling of two Lorenz systems as in drive-response mode given by  
\begin{eqnarray}
    {{dx_1} \over {dt}} & = & \sigma (x_2 - x_1) + \epsilon (y_1 - x_1)    \nonumber  \\ 
    {{dx_2} \over {dt}} & = & \gamma x_1 - x_1x_3 - x_2                          \nonumber  \\ 
    {{dx_3} \over {dt}} & = & x_1x_2 - \rho x_3                                \nonumber  \\
    {{dy_1} \over {dt}} & = & \tau_m (\sigma(y_2 - y_1))                   \nonumber  \\
    {{dy_2} \over {dt}} & = & \tau_m(\gamma y_1 - y_1y_3 - y_2)             \nonumber  \\  
    {{dy_3} \over {dt}} & = & \tau_m(y_1y_2 - \rho y_3) 
    \label{eq:8}
\end{eqnarray}     
with the parameters in the chaotic regime as $\sigma = 10, \gamma = 28, \rho = 8/3$.  
With $\epsilon = 0.05$, the control parameter 
$\tau_m$ is varied and for each $\tau_m$, $2 \times 10^5$ trajectory points are used for 
computing the scaling region using the modified box counting code after discarding the 
first $20000$ points as transients. 

\begin{figure}
\begin{center}
\includegraphics*[width=16cm]{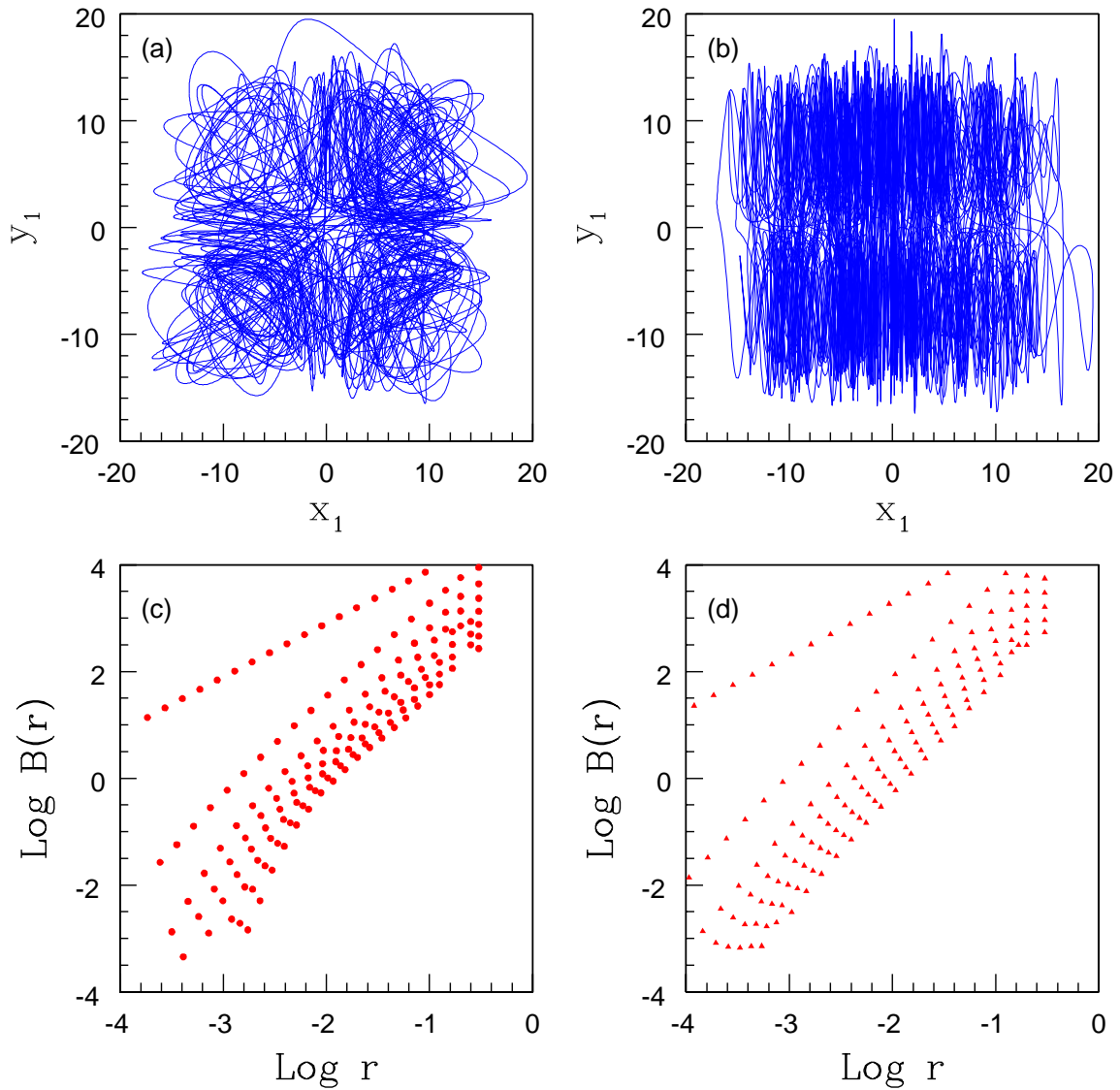}
\end{center}
\caption{Top panel shows the attractors obtained by drive-response coupling two Lorenz 
systems, with $\epsilon = 0.05$, as in Eq. (6) for (a) $\tau = 2.4$ and (b) $\tau = 20.0$. 
The corresponding scaling 
regions for the weighted box counting sum $B(r)$ from $y_1$ time series are shown in the lower 
panels (c) and (d) respectively. The change in the scaling region is obvious for the two $\tau$ values. 
While the left one is a four wing hyperchaotic attractor, the one on the right is chaotic.}
\label{f.8}
\end{figure}

Since we already take the time step of integration to be very small, if $\tau_m << 1$, 
the evolution of the second system turns out to be very slow and the effect of coupling 
will be a very small perturbation on the first system. In effect, the resulting attractor 
is found to be chaotic with $D_2$ close to that of a single system. On the other hand, 
when $\tau_m >> 1$, the second system evolves very fast and often swamps out the dynamics 
arising out of the evolution of the first system. The result is again a chaotic attractor, 
but with dimension higher than the individual systems $(D_2 > 3)$, resulting in 
high dimensional chaos. 
In between, for a range of values of $\tau_m$, the resulting attractor is found to 
display hyperchaotic behavior. In Fig.~\ref{f.8}, we show the results for two values of 
$\tau_m$, namely, $2.4$ and $20.0$, the former being hyperchaotic and the latter chaotic 
as reflected by the change in the scaling regions for the two cases. Note that this is 
analogous to what we have found in \S 4 in the case of Cantor set where an intermediate 
range of iteration steps $m$ leads to a Cantor set with dual slopes.  

The coupled Lorenz model with two different time scales as we have considered, but with 
two way coupling, has been used earlier as ocean-atmospheric model in climate studies 
\cite {siqu,sold} apart from synchronization studies \cite {kajari}. 
The model represents the interactive dynamics of a fast changing 
atmosphere and slow fluctuating ocean. We have also analysed this model numerically 
with $\epsilon = 0.05$ for a range of values of $\tau_m$. Our results indicate that 
the underlying dynamics can be hyperchaotic or high dimensional chaotic 
depending on the value of the time scale parameter $\tau_m$. 

\begin{figure}
\begin{center}
\includegraphics*[width=16cm]{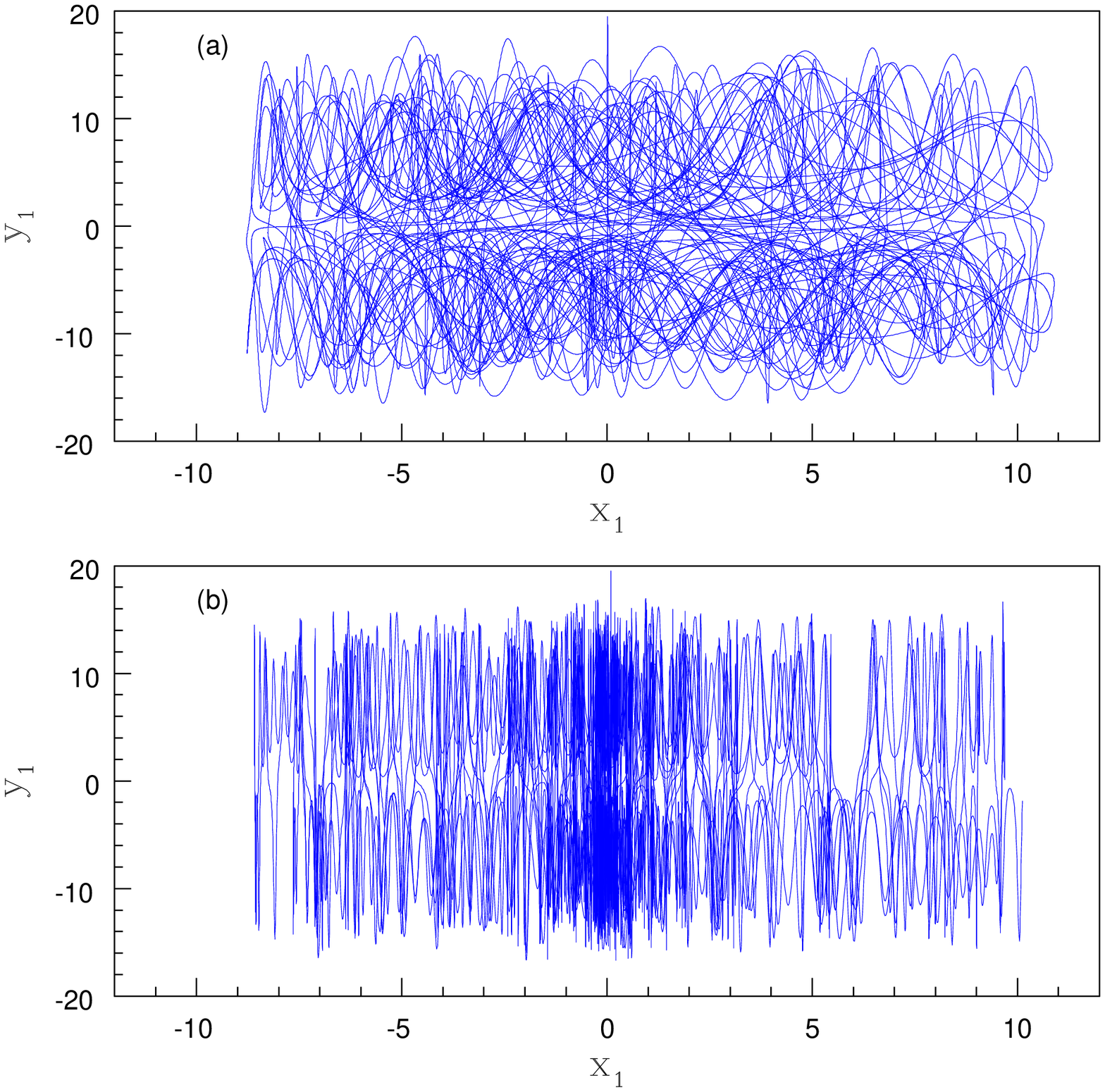}
\end{center}
\caption{Top panel is a hyperchaotic attractor arising out of the mutual diffusive coupling 
between the Lorenz and R\"ossler attractor for $\tau_m = 1.2$ and the bottom panel is a chaotic attractor 
for the same coupling with $\tau_m = 12.4$. In both cases, value of $\epsilon = 0.05$.}
\label{f.9}
\end{figure}

The second case we show is a mutual diffusive coupling between the standard Lorenz and R\"ossler 
chaotic attractors given by
\begin{eqnarray}
    {{dx_1} \over {dt}} & = & - (x_2 + x_3) + \epsilon (y_1 - x_1)              \nonumber  \\ 
    {{dx_2} \over {dt}} & = & x_1 + ax_2                                        \nonumber  \\ 
    {{dx_3} \over {dt}} & = & b + x_3(x_1 - c)                                   \nonumber  \\
    {{dy_1} \over {dt}} & = & \tau_m (\sigma(y_2 - y_1)) + \epsilon (y_1 - x_1)   \nonumber  \\
    {{dy_2} \over {dt}} & = & \tau_m(\gamma y_1 - y_1y_3 - y_2)                    \nonumber  \\  
    {{dy_3} \over {dt}} & = & \tau_m(y_1y_2 - \rho y_3) 
    \label{eq:9}
\end{eqnarray}  
with $a = b = 0.2$ and $c = 7.8$. Here again we find that the resulting attractor is 
hyperchaotic for a range of intermediate values of $\tau_m$ and two typical cases, 
one hyperchaotic and the other high dimensional chaotic, are 
shown in Fig.~\ref{f.9}. Thus we find that any two regular chaotic attractors can be 
coupled to generate hyperchaos as well as high dimensional chaos by varying the 
value of $\tau_m$.

\section{Discussion and Conclusion}
Multifractality exhibited in multiscales has become an important tool for the analysis of 
complex systems. In this paper, we show numerically that the geometric structure of a 
hyperchaotic attractor is equivalent to that obtained by a superposition of two 
multifractal sets, with two distinct $f(\alpha)$ spectra superposed.  
This is shown explicitly by computing the $f(\alpha)$ spectra for two different 
class of hyperchaotic attractors. We show that the hallmark of such a structure is a 
cross-over behavior in the scaling region of correlation dimension. Based on our numerical 
results, we conjecture that a hyperchaotic attractor is a superposition of two multifractals. 

Hyperchaos is normally characterized by computing the 
spectra of LEs and looking at the transition of the second largest LE above zero. Though 
this method works for synthetic systems, it becomes much more difficult when the system 
is represented as a time series. Our numerical results indicate that there is also a 
structural change for the underlying attractor as the system makes a transition to 
hyperchaos. This structural change can be more easily identified using the spectrum of 
dimensions, especially for systems analysed using time series.  Thus, our results also 
offer a method to identify transition to hyperchaos through correlation dimension 
analysis of time series data. As another application, we present a general scheme by coupling 
two chaotic attractors to get chaos, hyperchaos and high dimensional chaos by varying a 
time scale parameter.  

There are two specific aspects we wish to highlight regarding the present work. We are aware 
that the topic of multifractality is intrinsically mathematical in nature and our claims here  
are based only on numerical results. A firm theoretical analysis is required to back up the results of 
numerical simulations presented here. We do hope that the numerical results are compelling 
enough to motivate a mathematical analysis in order to establish the results presented here. 
Secondly, we address the question how the information that hyperchaos involves a dual 
multifractal spectrum is important from a practical point of view. Though many authors have 
produced hyperchaos by coupling two low dimensional chaotic attractors, why and under what 
conditions this can be achieved has not been clear so far. Our results clarify both these 
questions and indicate that hyperchaos can be generated by suitably coupling any two low 
dimensional chaotic systems. Moreover, by tuning a time-scale parameter, it is possible to 
generate not only hyperchaos, but high dimensional chaos as well. This could be especially 
important in many practical applications, such as, cryptography and secure communication 
where the time series carrying encripted messages needs to be novel and sufficiently complex. 

Just like the dual positive LEs, a dual multifractal spectrum also appears to be characteristic 
of every hyperchaotic attractor as per our numerical results. However, such a structure is not 
unique to hyperchaotic attractors. We explicitly 
show an example where such a structure can be synthetically generated from a Cantor set. 
To our knowledge, the 
concept of a superposition of two multifractal sets to characterize the structure of a fractal 
object is novel and has not been discussed in the literature 
either in the context of real world fractals or those generated from dynamical systems. The 
results presented here clearly show that the fractal structure of a hyperchaotic attractor 
is qualitatively different from that of a chaotic attractor and may serve as a first step 
towards a better understanding of the highly complex structure of hyperchaotic attractors 
in high dimensional systems.

{\bf Acknowledgments} 

The authors thank R. E. Amritkar for the idea of generating dual multifractal from a 
Cantor set. KPH  acknowledges the hospitality and computing facilities in IUCAA, Pune. 

\bibliographystyle{elsarticle-num}

\end{document}